\documentclass[a4paper]{article}
\normalsize
\begin{document}
\title{Dyon Solution in Einstein-Yang-Mills Theory on a Cylindrical
Symmetric Space time with Cosmological Constant }
\author{{\bf R. J. Slagter}\footnote{E-mail:
{\tt info@asfyon.nl}}
 \\ University of Amsterdam, Physics Department, The Netherlands \\ and \\ 
 ASFYON, Astronomisch Fysisch Onderzoek Nederland }
\maketitle
\begin{abstract}
We investigated numerically dyon-like solutions of the SU(2) Einstein-Yang-Mills system
on a cylindrically symmetric space time with a cosmological constant. We find a new kind of
behaviour not found in the spherically symmetric models. For positive values of
$\Lambda$ we have an oscillatory behaviour of the magnetic component of the YM field around the r-axis,
so there is an arbitrary number of nodes. For increasing positive $\Lambda$,
the frequency increases also and the solution breaks down at finite $r=r_h$, indicating a singularity. 
The electric component, however, approaches a constant value. After further increasing $\Lambda$,
this global behaviour repeats itself at a larger $r$ while the former singular behaviour disappears. This
behaviour depends critically on the position of the core of the string.
For increasing negative $\Lambda$, the oscillatory behaviour disappears and the magnetic and 
electric components behave like the scalar and gauge field in the Abelian cosmic string model.

\end{abstract}

\section{Introduction}
It is known for decades that the static Einstein-Yang-Mills theory for the gauge group SU(2) admits regular 
solutions. The first example was that of Bartnik and McKinnon (BK)~\cite{Bartnik}. The basic idee behind this phenomenon 
is the balance between the attractive gravity and the repulsive Yang-Mills field. When a cosmological constant is 
incorporated in the model, the behaviour of the particle-like solution can be significant different compared to the
original BK solution~\cite{Bjoraker,Breit,Volkov,Torri}. For a nice overview, see~\cite{Volkov2}.

When the cosmological constant is negative, there are two types of generic solutions, the singular ones and 
the asymptotical anti-de Sitter (AdS) solutions.
There is also a change of the topology of  the moduli space of globally regular ('soliton') and black hole solutions
when going from negative to positive values of $\Lambda$. This is  called a 'fractal' structure~\cite{Breit}. 
 Moreover, it is conjectured that negative $\Lambda$ solutions are the only stable EYM solutions and the magnetic
component of the YM field need not be nodeless in order to be stable. In these models, one usually takes
the pure magnetic ansatz.
Further, one can investigate the dyon-type solutions, where  a non-zero electric component of the 
YM field is incorporated~\cite{Bjoraker}.
It turns out that dyon solution can exist in these models with negative $\Lambda$, however under quite
special asymptotic assumptions.
In the present paper we investigate the EYM system with  positive and negative cosmological constant
on a cylindrical space time, without any assumption on the
asymptotic behaviour and try to find  a connection between the Abelian sting-like solution found on the same space time.


\section{The field equations of the Einstein Yang-Mills system}

Consider the Lagrangian of the SU(2) EYM system
\begin{equation}
S=\int d^4x\sqrt{-g}\Bigl[\frac{1}{16\pi G}({\cal R}-2\Lambda )-
\frac{1}{4}{\cal F}_{\mu\nu}^a{\cal F}^{\mu\nu a}\Bigr],
\end{equation}
with the YM field strength
\begin{equation}
{\cal F}_{\mu\nu}^a =\partial_\mu A_\nu^a-\partial_\nu
A_\mu^a+g\epsilon^{abc}A_\mu^bA_\nu^c,
\end{equation}
g the gauge coupling constant, G Newton's constant, $A_\mu^a$ the gauge potential,
and ${\cal R}$ the curvature scalar. The field equations then become
\begin{equation}
G_{\mu\nu}=8\pi G {\cal T}_{\mu\nu}-\Lambda g_{\mu\nu},
\end{equation}
\begin{equation}
{\cal D}_\mu {\cal F}^{\mu\nu a}=0,
\end{equation}
with ${\cal T}$ the energy-momentum tensor
\begin{equation}
{\cal T}_{\mu\nu}={\cal F}_{\mu\lambda}^a{\cal F}_\nu^{\lambda a}
-\frac{1}{4}g_{\mu\nu}{\cal F}_{\alpha\beta}^a {\cal F}^{\alpha\beta a},
\end{equation}
and ${\cal D}$ the gauge-covariant derivative,
\begin{equation}
{\cal D}_\alpha{\cal F}
_{\mu\nu}^a\equiv \nabla_\alpha{\cal F}_{\mu\nu}^a+g\epsilon^{abc}A_\alpha^b{\cal F}
_{\mu\nu}^c.
\end{equation}
Let us consider the stationairy exterior line element
\begin{equation}
ds^2=C(-dt^2+dz^2)+dr^2+B^2r^2 d\varphi^2,
\end{equation}
where $B$ and $C$ are functions of r.
For the YM-field we will take~\cite{Slagter1,Slagter2}
\begin{eqnarray}
A_\mu dx^\mu =(\Phi_1\hat\tau_r+\Phi_2\hat\tau_z)dt + A_1\hat\tau_\varphi dr +A_2\hat\tau_\varphi dz 
+(W_1\hat\tau_r+(W_2-1)\hat\tau_z)d\varphi
\end{eqnarray}
with $\Phi_i, A_i$ and $W_i$ functions of r, and $\hat\tau_i$ the orthonormal axial generators of the SU(2) 
\begin{eqnarray}
\hat\tau^{(n)}_r&=&(cos(n\varphi),sin(n\varphi),0), \cr
\hat\tau^{(n)}_z&=&(0,0,1), \cr
\hat\tau^{(n)}_\varphi&=&(sin(n\phi),-cos(n\phi),0) ,
\end{eqnarray}
with n the magnetic charge.
For the special case, $A_1=A_2=0$, $\Phi_2=W_1=0$ and n=1, we obtain a consistent set of
equations:
\begin{eqnarray}
C''&=&\frac{1}{4C}(C')^2-\frac{\kappa}{B^2 r^2}\Bigl[\Phi^2W^2-B^2 r^2(\Phi')^2-C(W')^2\Bigr]-
\lambda C ,\cr
B''&=&\frac{B}{4C^2}(C')^2-\frac{2}{r} B'+\frac{\kappa}{BCr^2}\Bigl[\Phi^2W^2
+B^2 r^2(\Phi')^2-2C(W')^2\Bigr], \cr
(rB\Phi')'&=&\frac{\Phi W^2}{Br}, \cr
(\frac{CW'}{Br})'&=&-\frac{\Phi^2 W}{Br},
\end{eqnarray}
where $\kappa\equiv\frac{4\pi G}{g^2}$.
The conservation law  $\nabla_\mu T^{\mu\nu}$ is also fulfilled.
Further, there is a scale invariance
\begin{equation}
C\rightarrow\gamma C,\quad \Phi\rightarrow \sqrt{\gamma}\Phi ,\quad B\rightarrow \beta B,\quad W\rightarrow \beta W,
\end{equation}
with $\beta$ and $\gamma$ some constant.

\section{Numerical solution}

The set of equations are easily solved numerically.
We used two different Fortran ODE-solvers to check the reliability of the numerical solution and the MAPLE-program
as final check. For all the routine-codes we find the same solution.
The absolute and relative tolerance was $1.10^{-15}$
For some specific values of  "shooting"parameters , we find a remarkable solution critically dependent
on the cosmological constant. 
Let us consider near the core of the string $r_0$ : 
\begin{equation}
W(r_0)=1-br_0^2 + ... , \qquad  \Phi(r_0)=ar_0 + ...
\end{equation}
Then we obtain for the initial values of the metric
components:
\begin{equation}
A(r_0)=1-\frac{1}{2}br_0^2 +{\cal O}(r_0^3), \qquad C(r_0)=1-\frac{a^2}{8b^2}\Bigl(\sqrt3\ln(3-
1.5\sqrt3)+6\Bigl)+{\cal O}(r_0^2)
\end{equation}
In the spherical symmetric situation (~\cite{Bjoraker}) the different solutions are
classified by some parameters determined by the condition of the field variables at  $r\rightarrow\infty$.
Here we are mainly interested in the critical situation for some initial values at $r_0$  
The energy density becomes
\begin{equation}
T_{tt}=\frac{A^2 r^2(\Phi')^2+\Phi^2 W^2+C (W')^2}{2g^2A^2r^2}.
\end{equation}
We observe that for the chosen initial values, $T_{tt}$ is finite for $r\rightarrow r_0$.

In figure 1 and 2 we plotted a typical solution for positive and negative values of $\Lambda$, and for $a=1, b=0.35$ and
for $r_0=0.01$.
For positive values of $\Lambda$ we observe an oscillatory behaviour of the magnetic component
around zero, so there are an arbitrary number of zero's. For increasing positive $\Lambda$,
the frequency increases also and the solution breaks down at finite $r=r_h$ where C approaches
zero. This means a cosmological horizon. However, for some value of $\Lambda$ ( for example $\Lambda$
around 0.004) the horizon is pushed to larger values of r, while the former horizon is resolved. 
This happens again for larger positive values of $\Lambda$.  In figure 3 we plotted the horizon $r_h$ as function of
$\Lambda$.
For increasing negative $\Lambda$ we observe that the oscillation of $W$ slows down and disappears at a critical
$\Lambda$. The solution behaves like a string solution. This fact was also observed in the radiative counterpart 
model~\cite{Slagter2}.
For smaller values of $r_0$ this  pushing foreward of $r_h$ dissapears. In figure 4 we plotted $r_h$ versus $\Lambda$
for $r_0=10^{-8}$.

\input epsf
\centerline
{\hfil\epsfysize=50mm\epsfbox{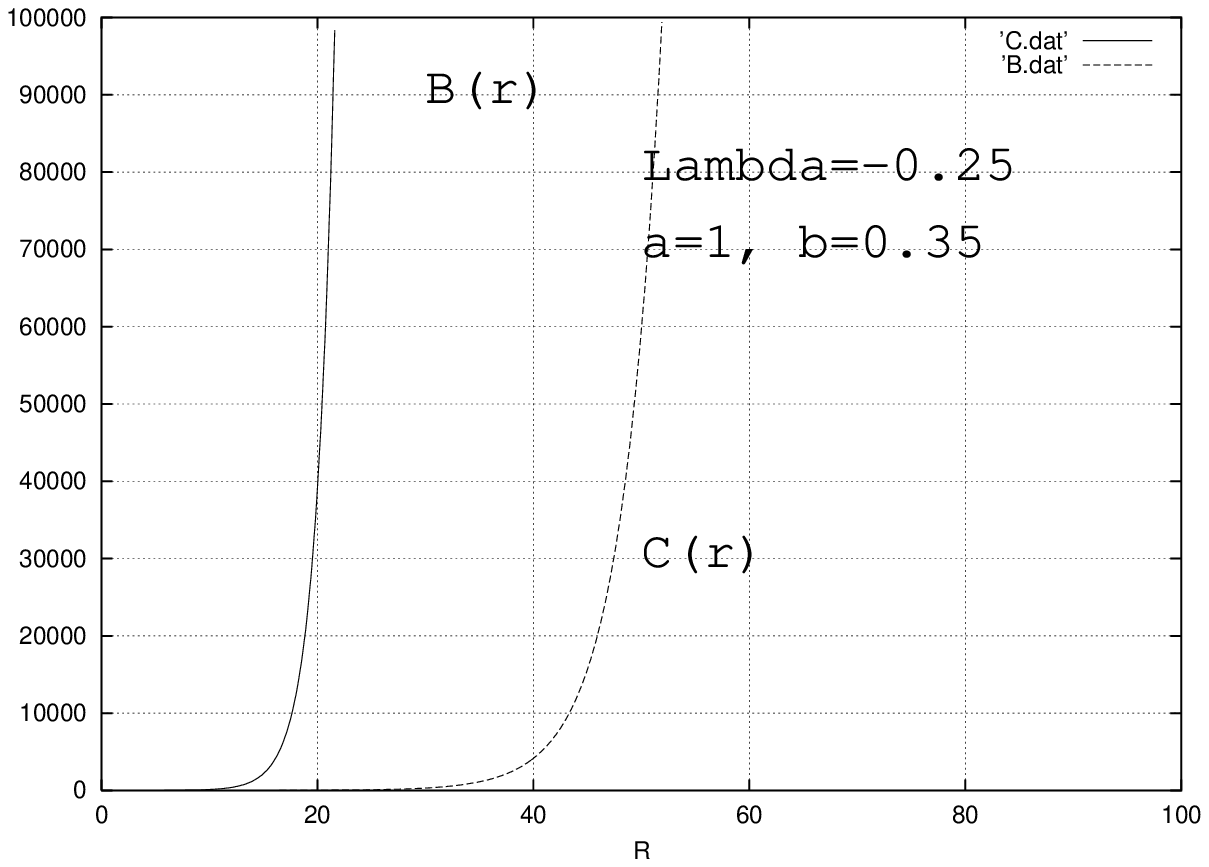}  
 \hfil\epsfysize=50mm\epsfbox{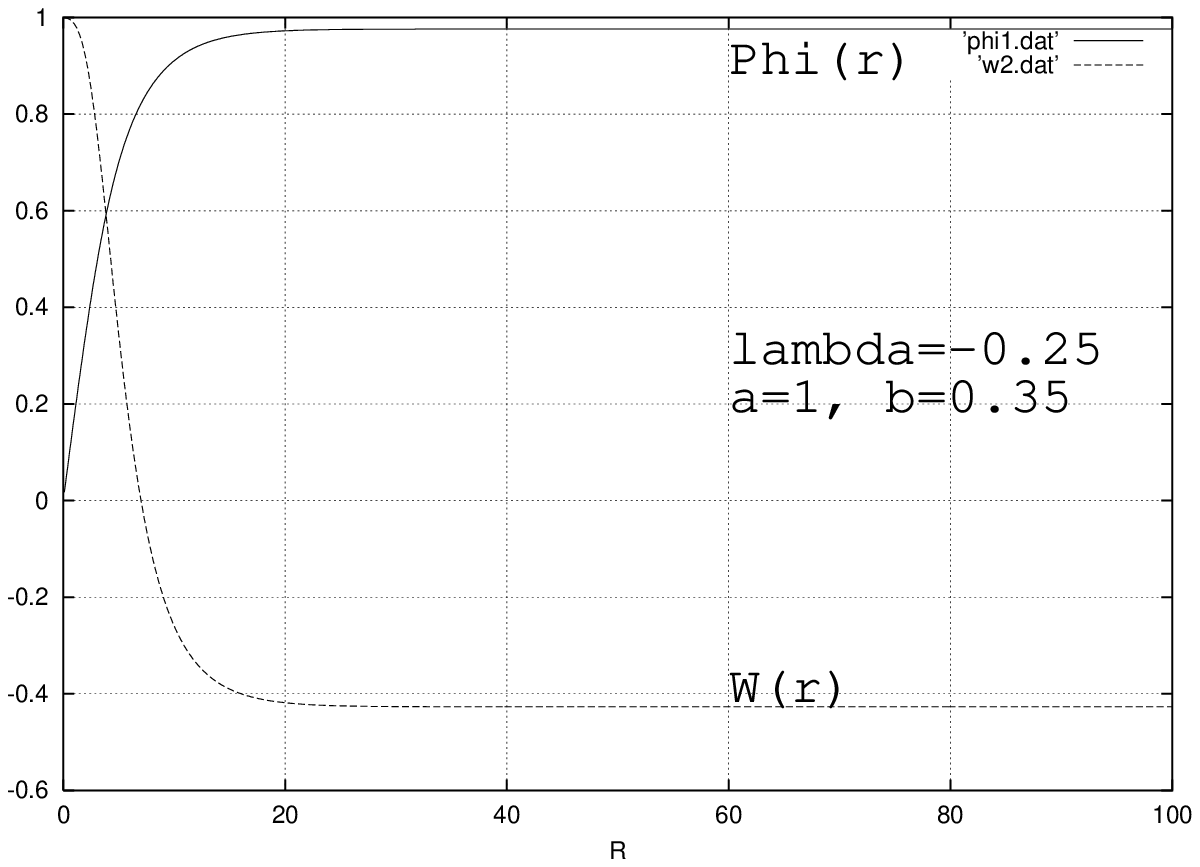}
\hfil}
\centerline
{\hfil\epsfysize=50mm\epsfbox{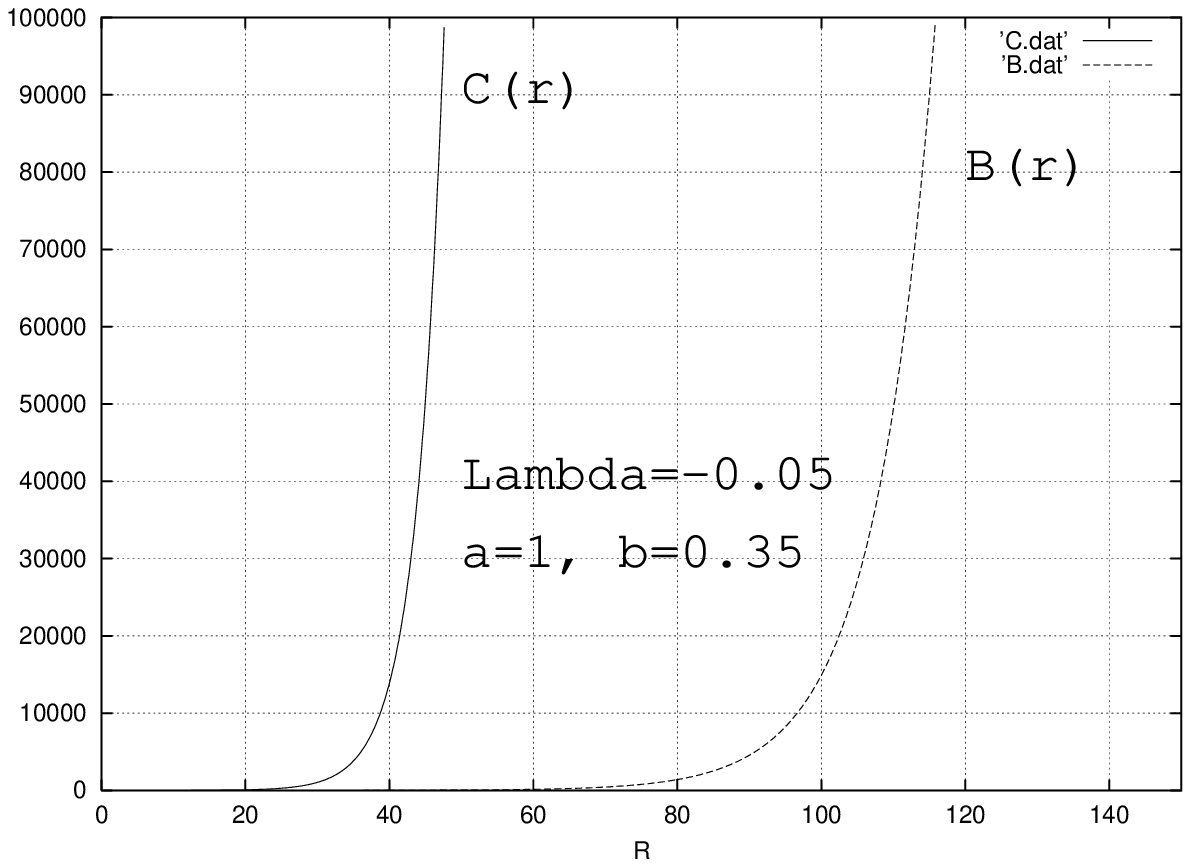}   
 \hfil\epsfysize=50mm\epsfbox{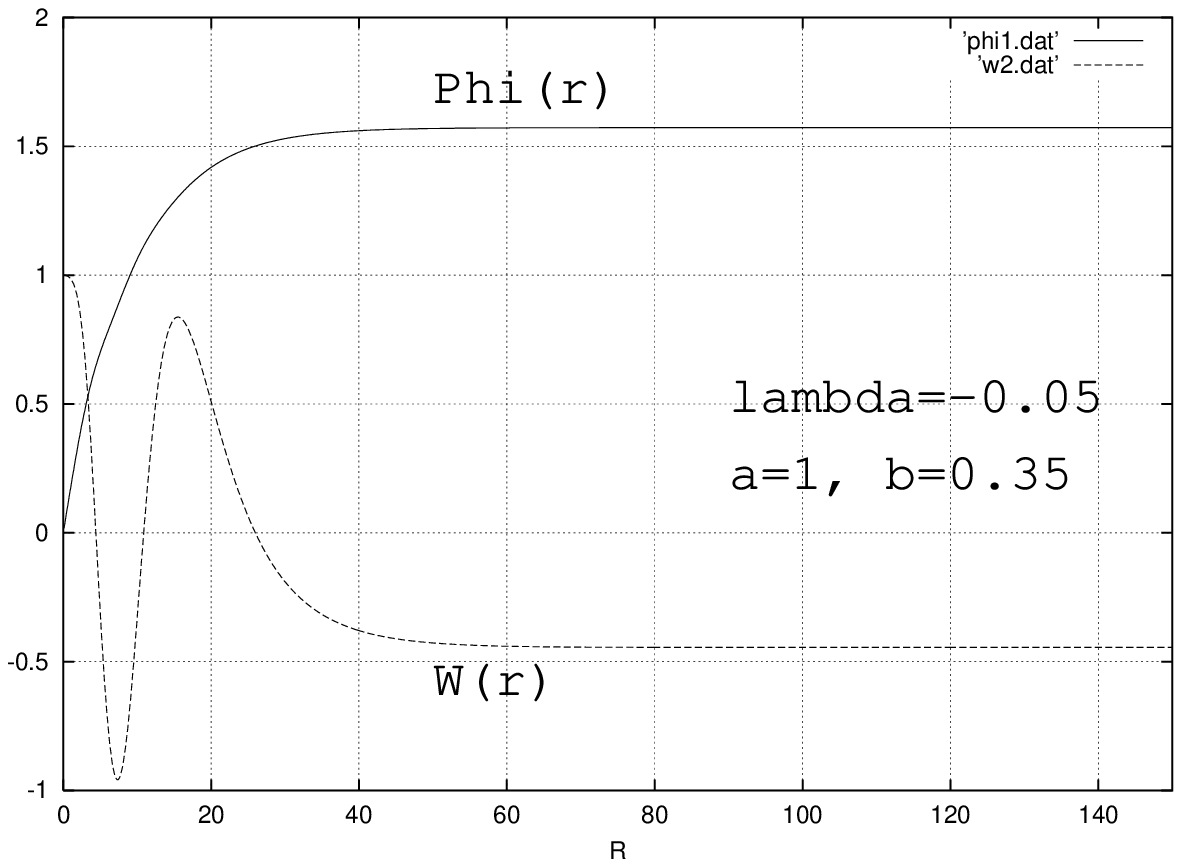}
\hfil}
\centerline
{\hfil\epsfysize=50mm\epsfbox{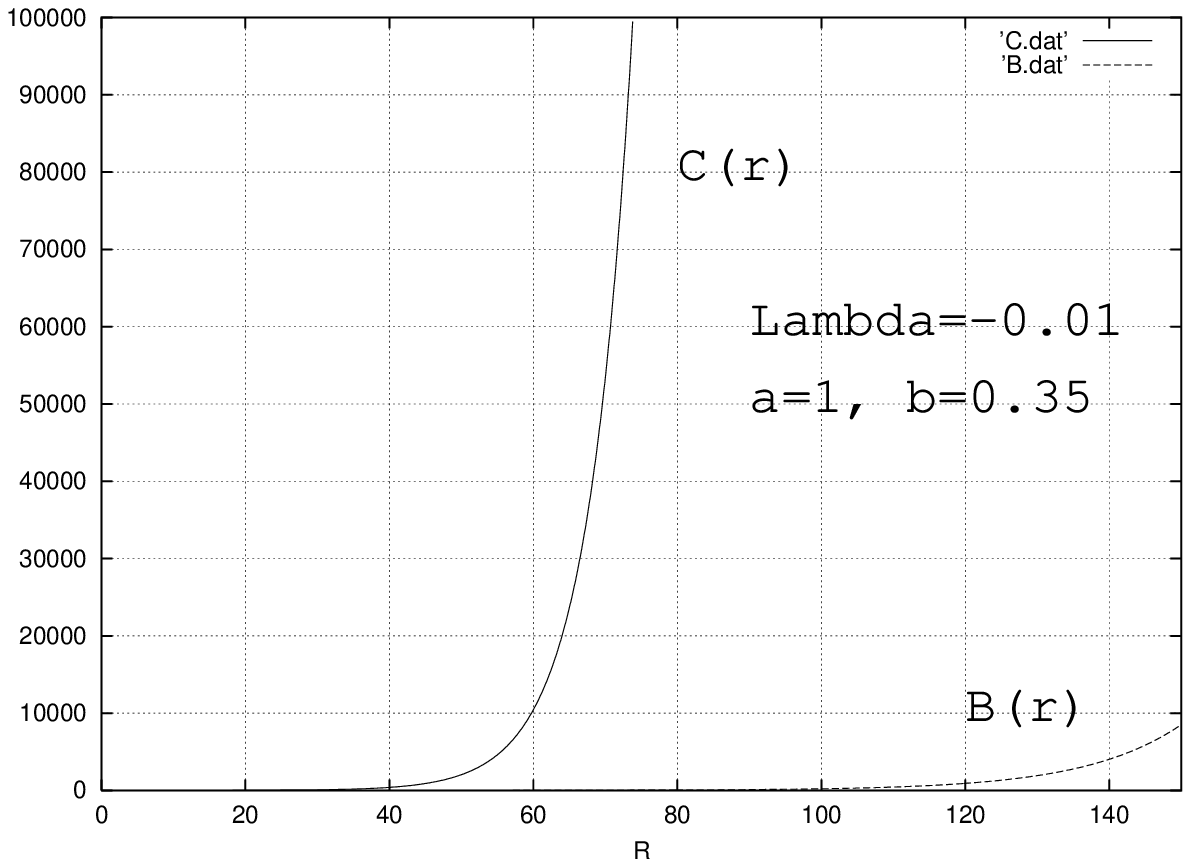}    
\hfil\epsfysize=50mm\epsfbox{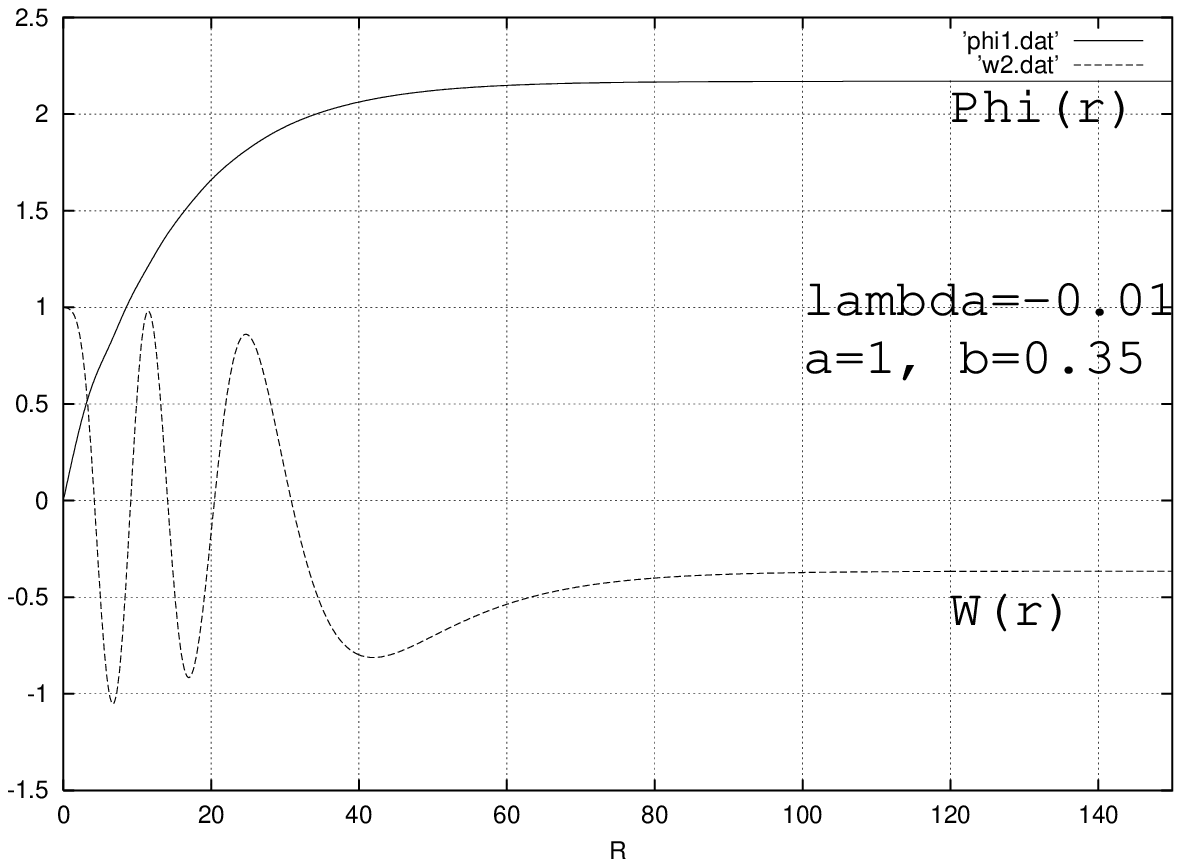}
\hfil}

Figure 1. Plot of $B(r), C(r), W(r)$, and $\Phi(r)$  for negative $\Lambda$. 
 We took for $\Lambda$ -0.25, -0.15, and  -0.01   respectively. $\kappa =1$ and $b=1, c=0.35$.$r_0=0.01$.
  

\centerline
{\hfil\epsfysize=50mm\epsfbox{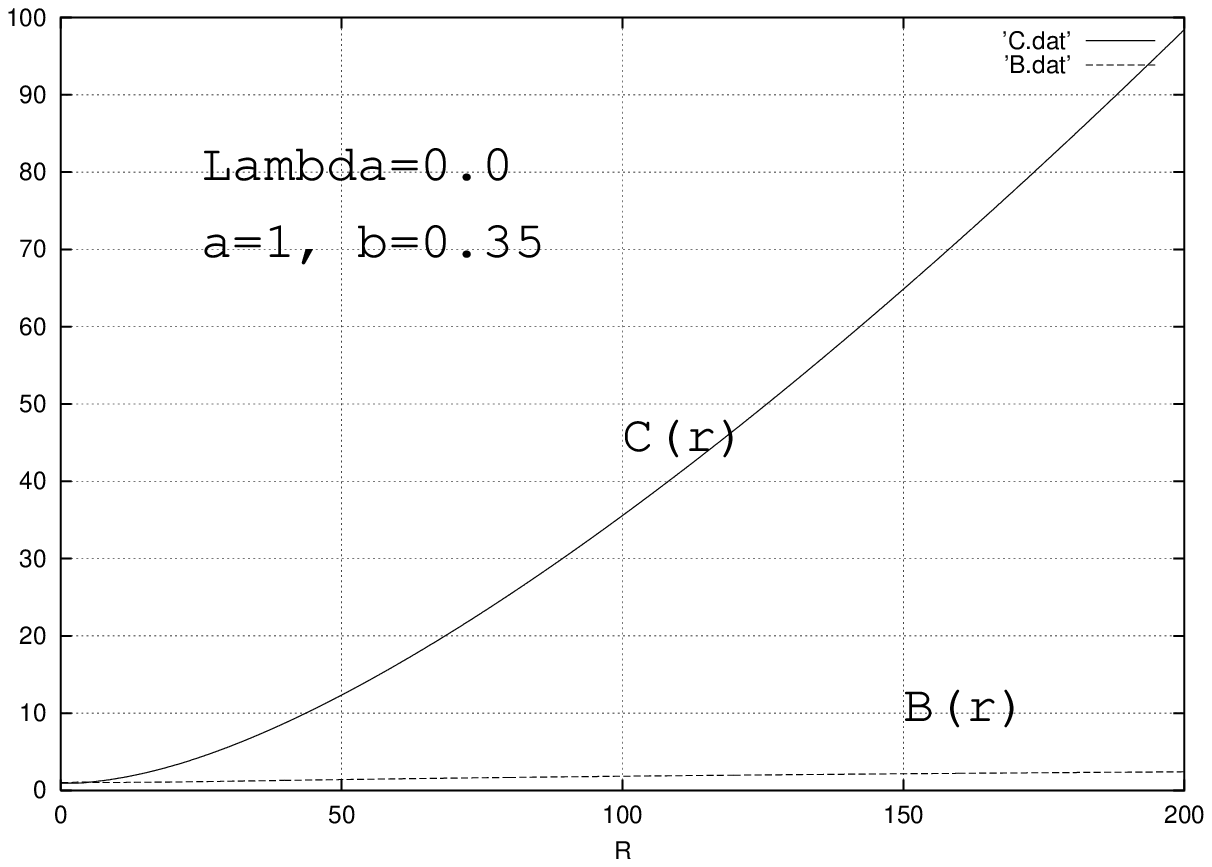}   
\hfil\epsfysize=50mm\epsfbox{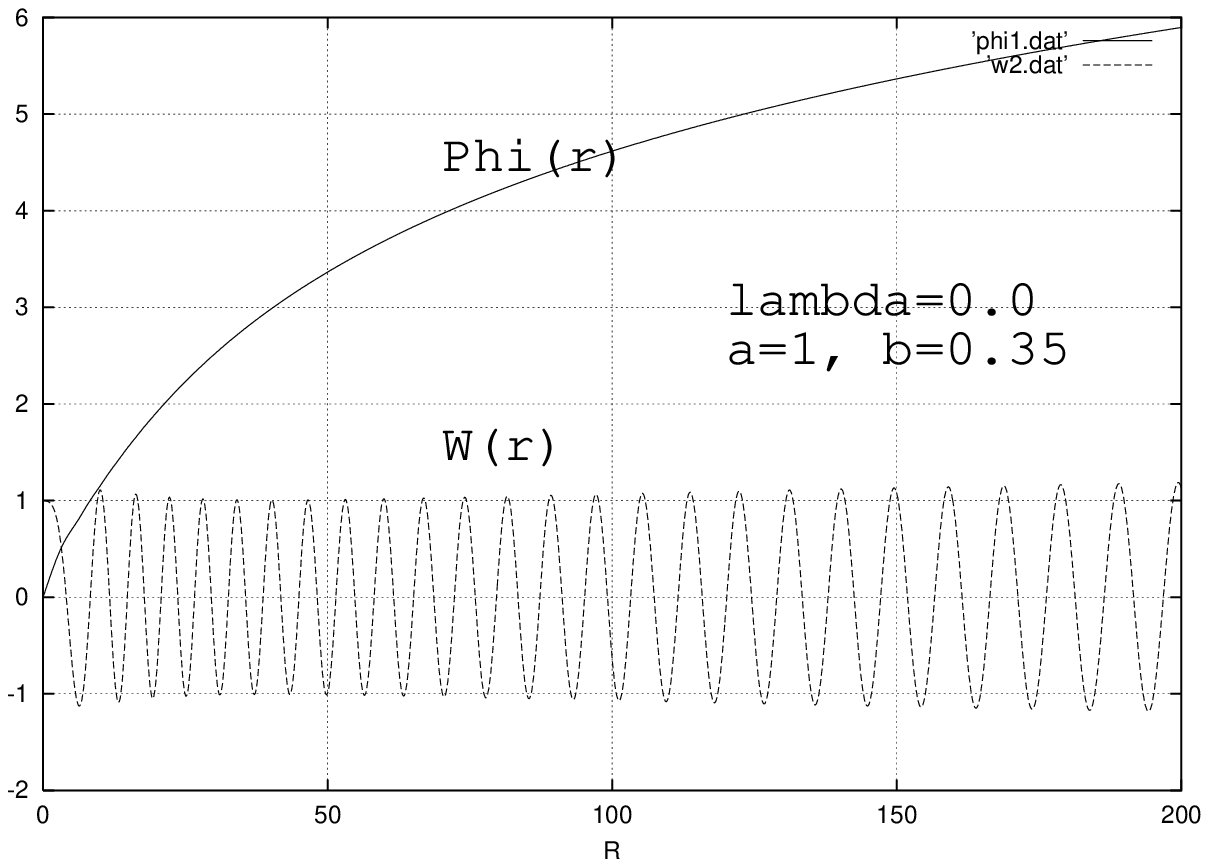}
\hfil}
\centerline
{\hfil\epsfysize=50mm\epsfbox{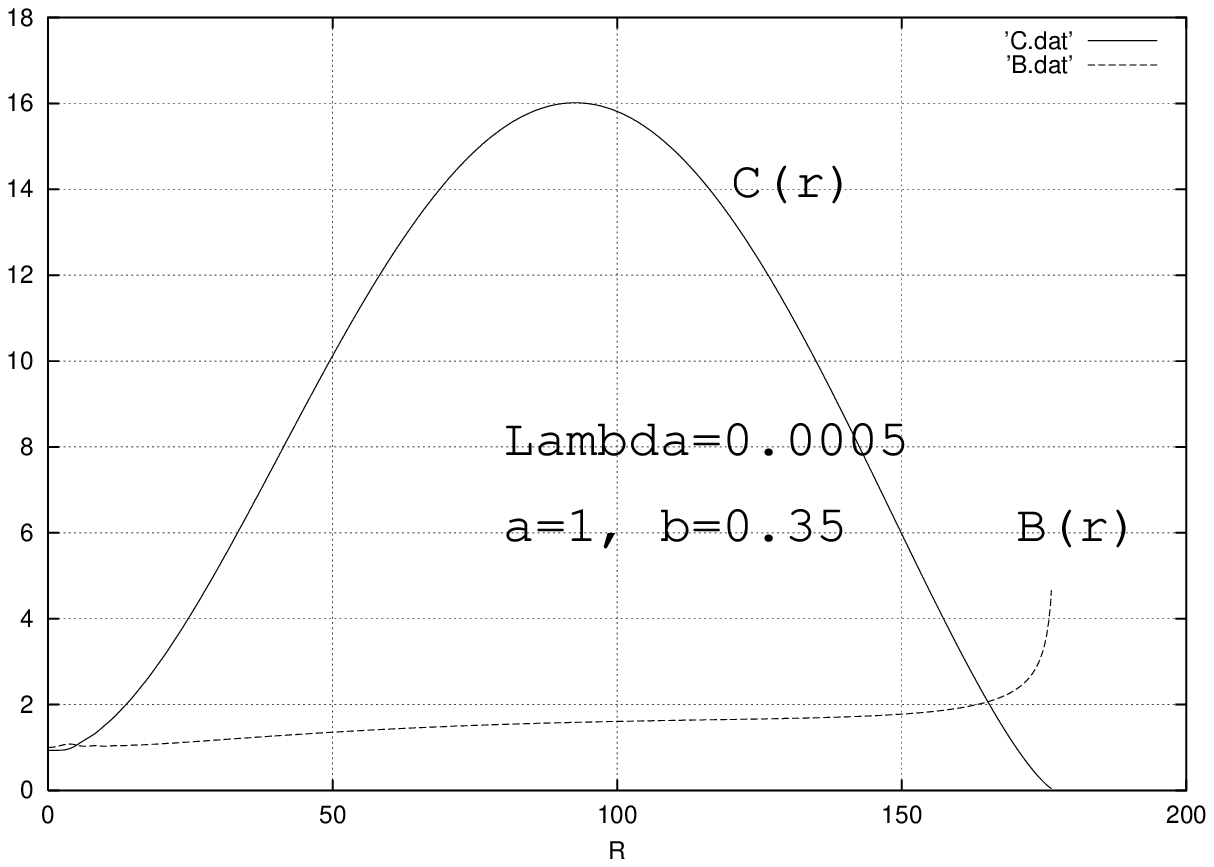}    
\hfil\epsfysize=50mm\epsfbox{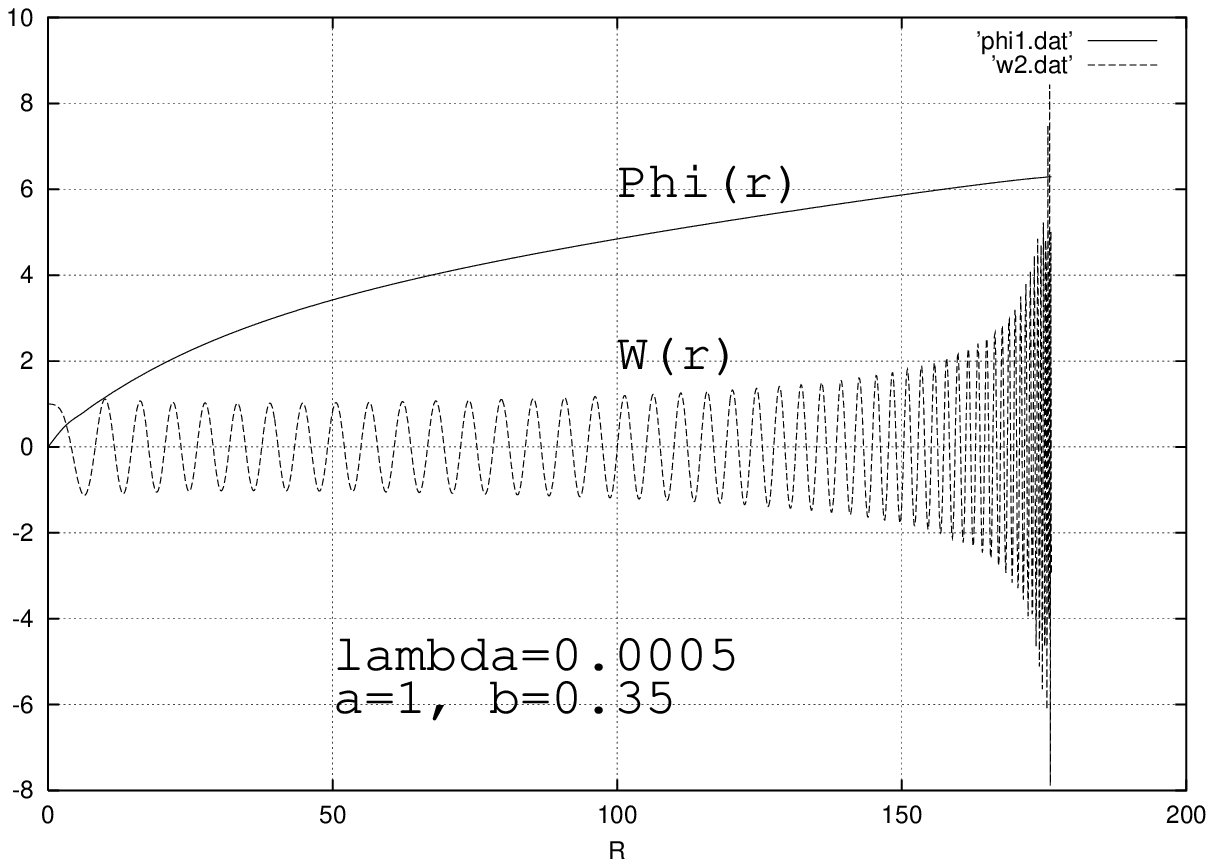}
\hfil}
\centerline
{\hfil\epsfysize=50mm\epsfbox{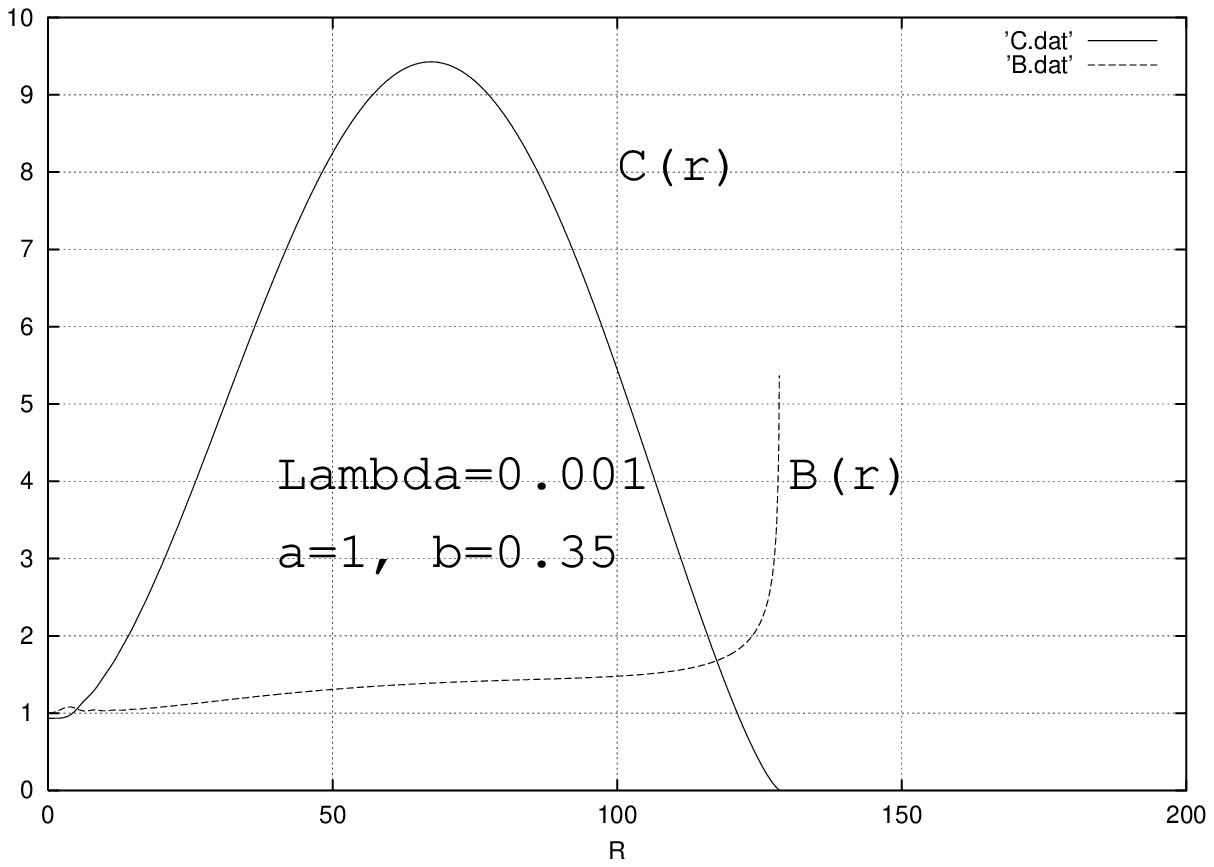}   
\hfil\epsfysize=50mm\epsfbox{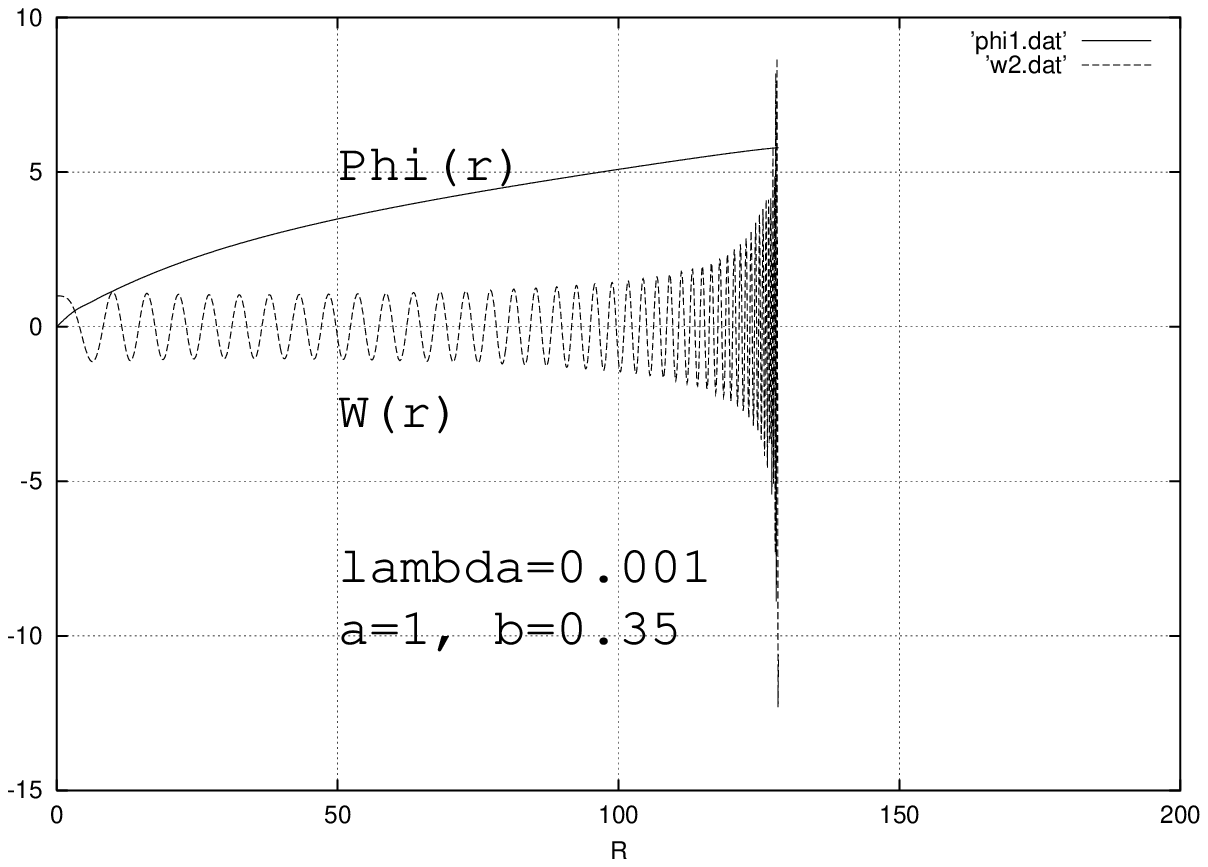}
\hfil}
\centerline
{\hfil\epsfysize=50mm\epsfbox{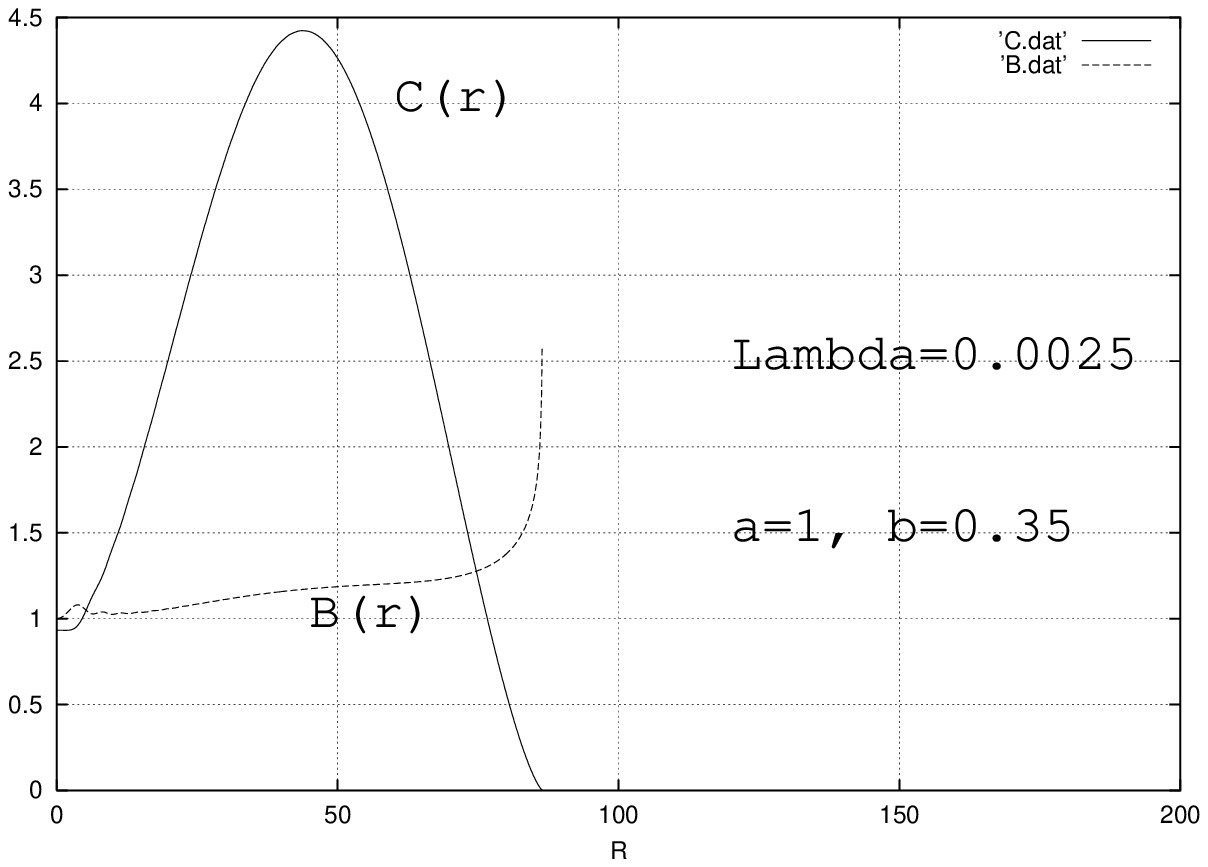}    
 \hfil\epsfysize=50mm\epsfbox{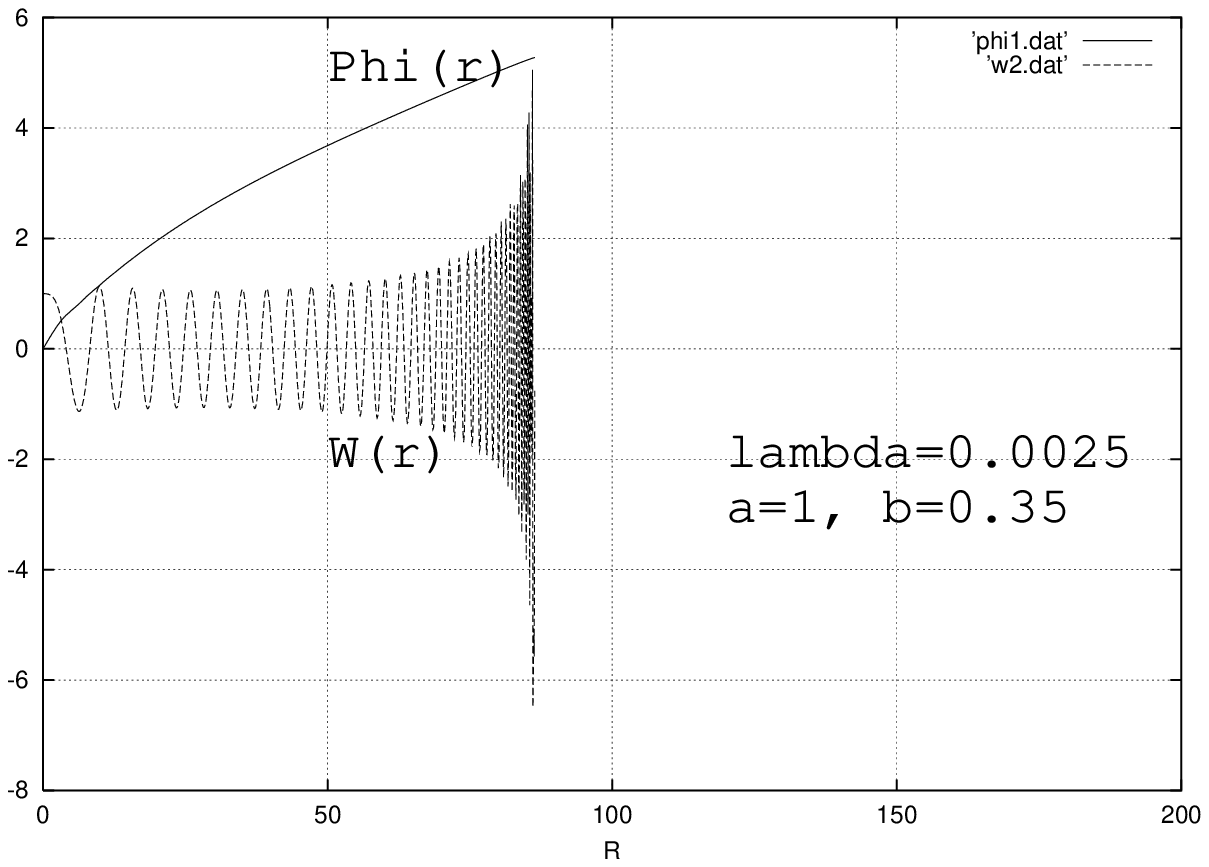}
\hfil}
\centerline
{\hfil\epsfysize=50mm\epsfbox{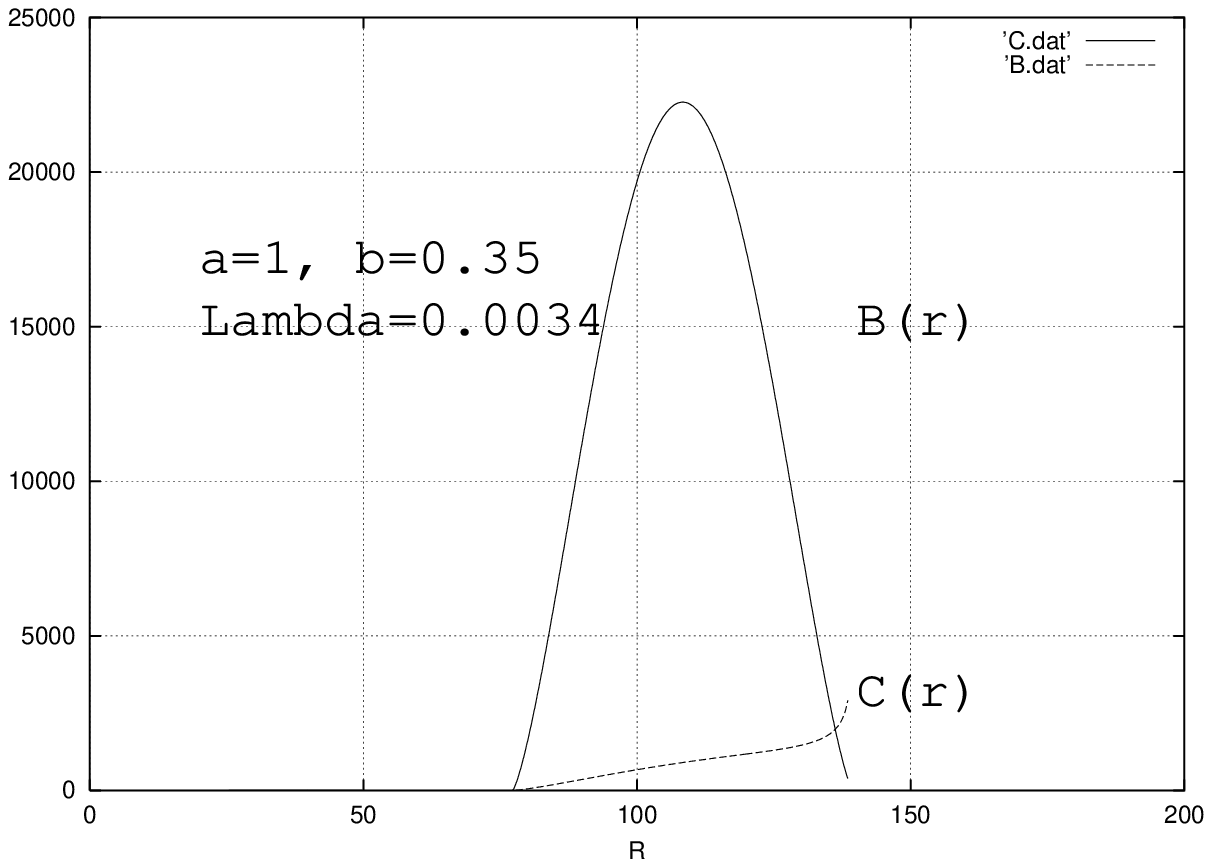}    
 \hfil\epsfysize=50mm\epsfbox{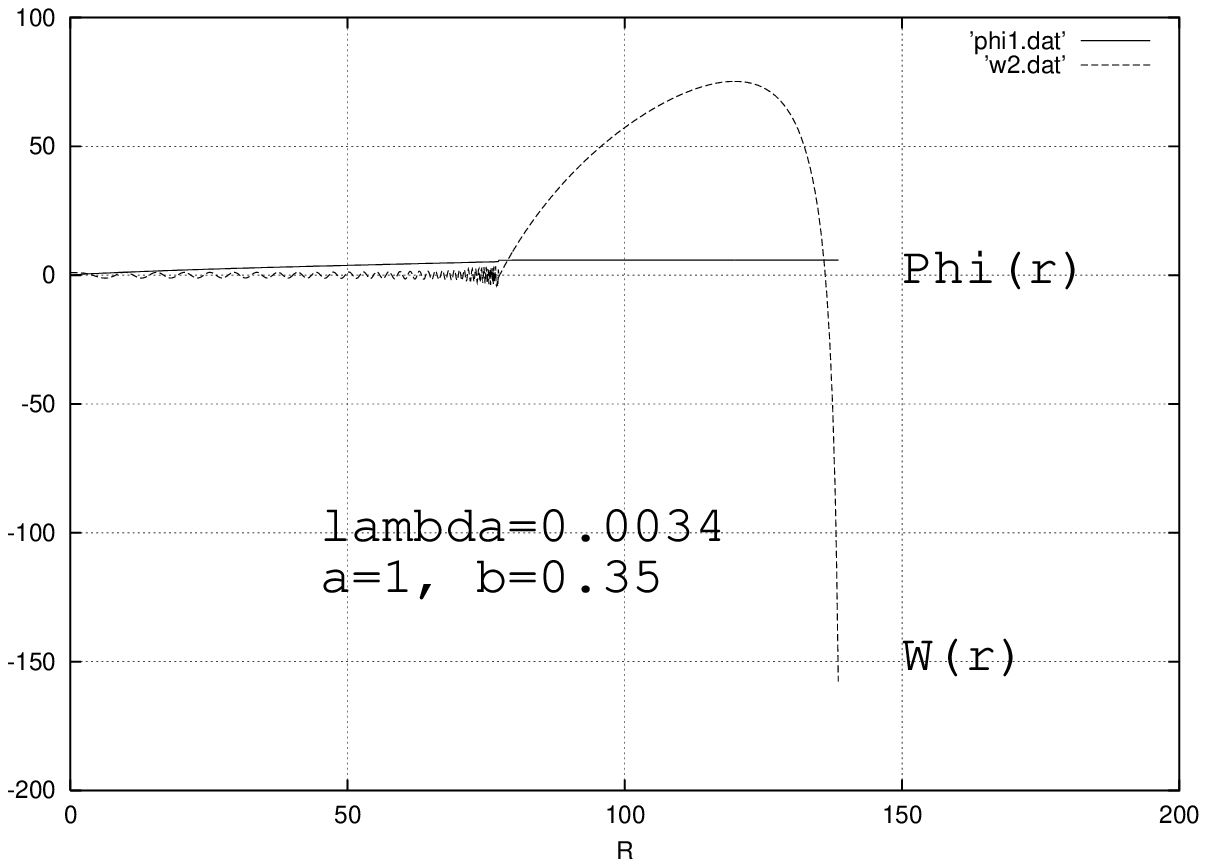}
\hfil}
\centerline
{\hfil\epsfysize=50mm\epsfbox{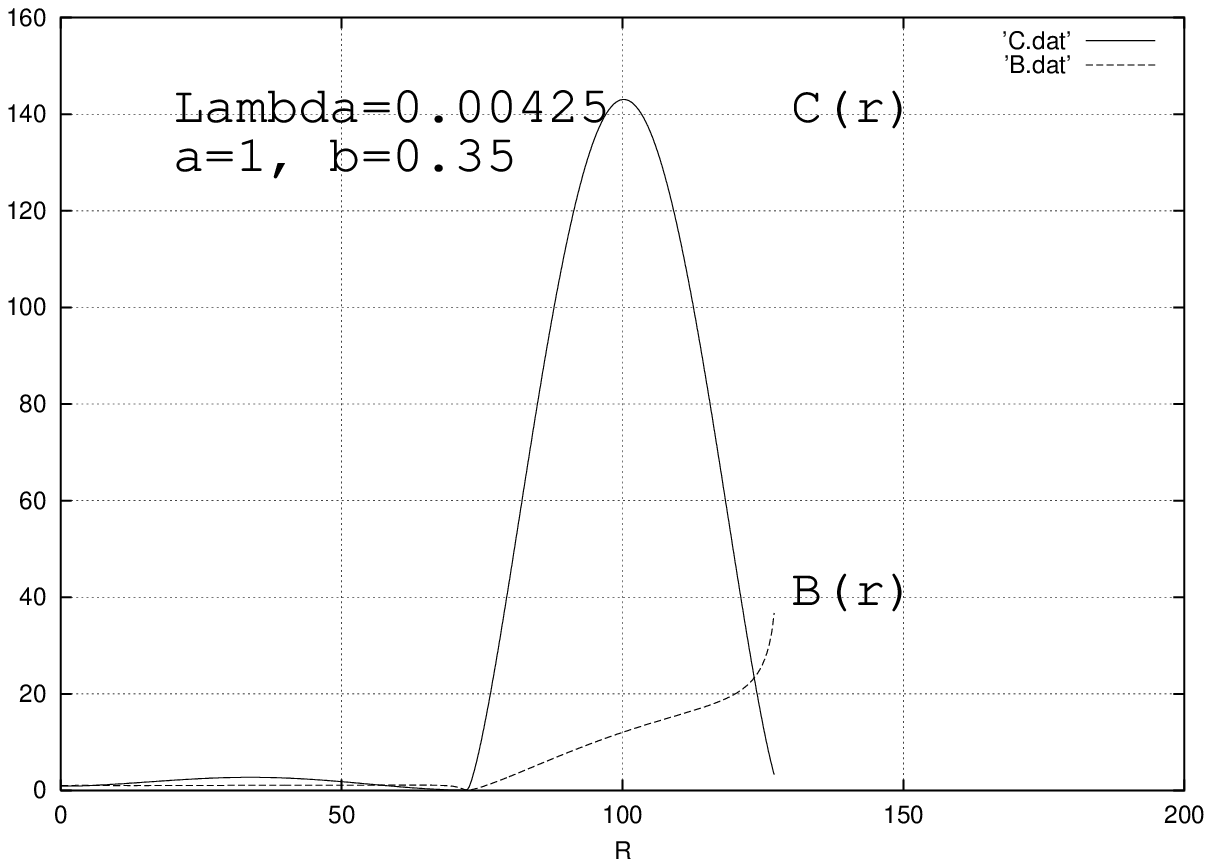}   
 \hfil\epsfysize=50mm\epsfbox{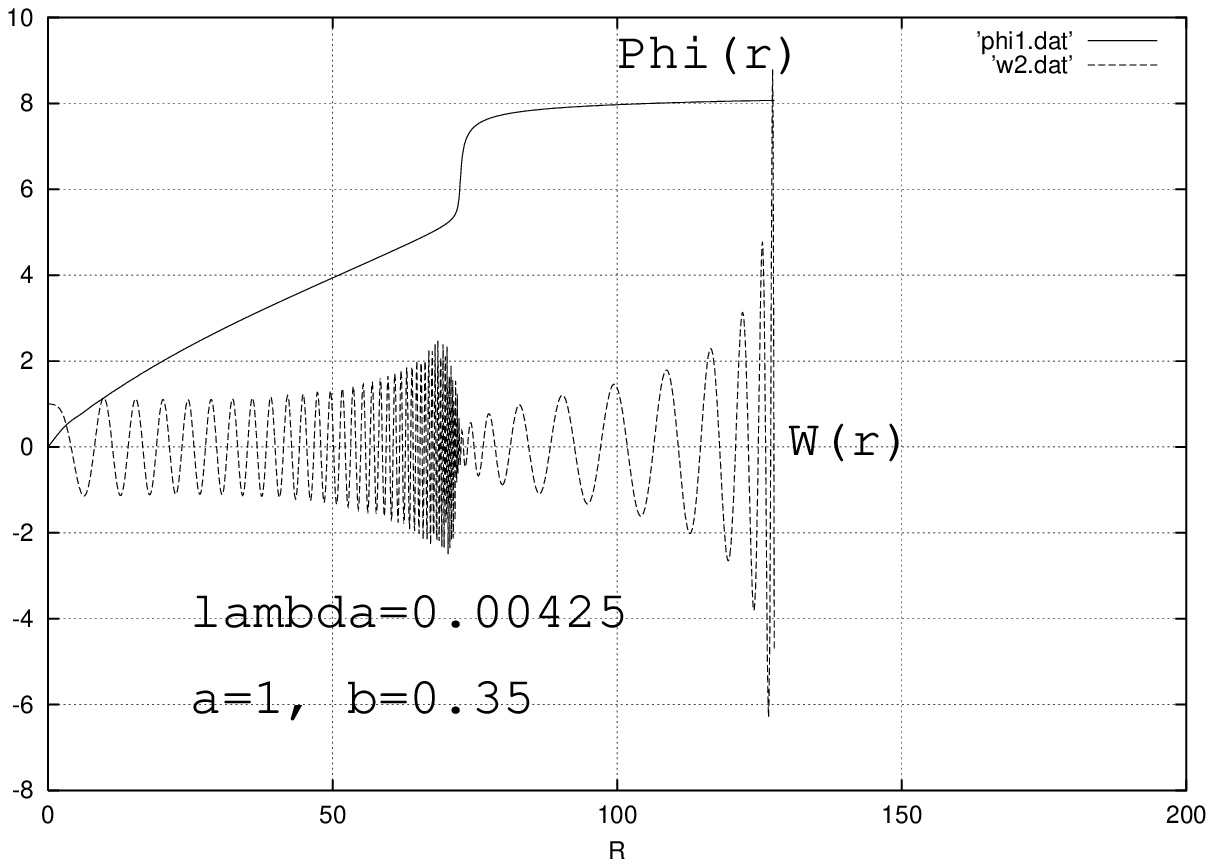}
\hfil}
\centerline
{\hfil\epsfysize=50mm\epsfbox{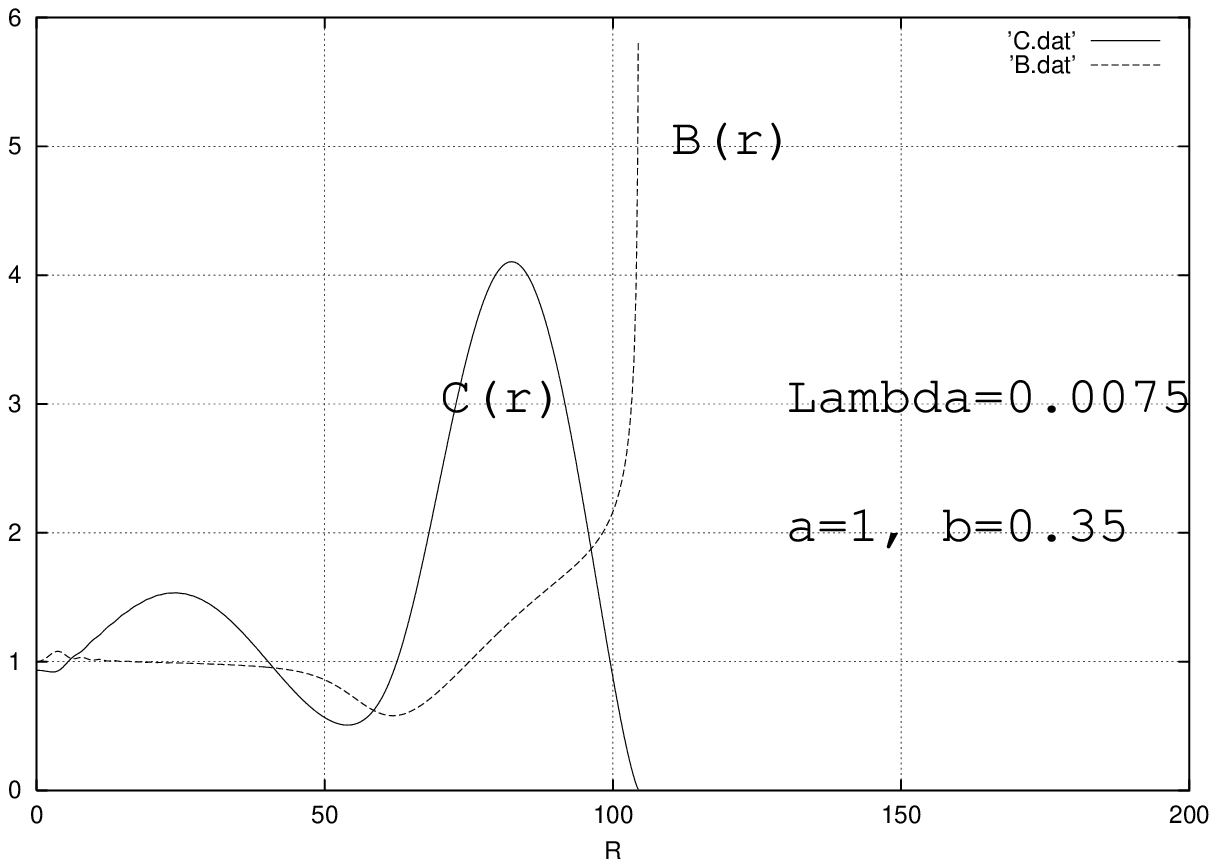}    
 \hfil\epsfysize=50mm\epsfbox{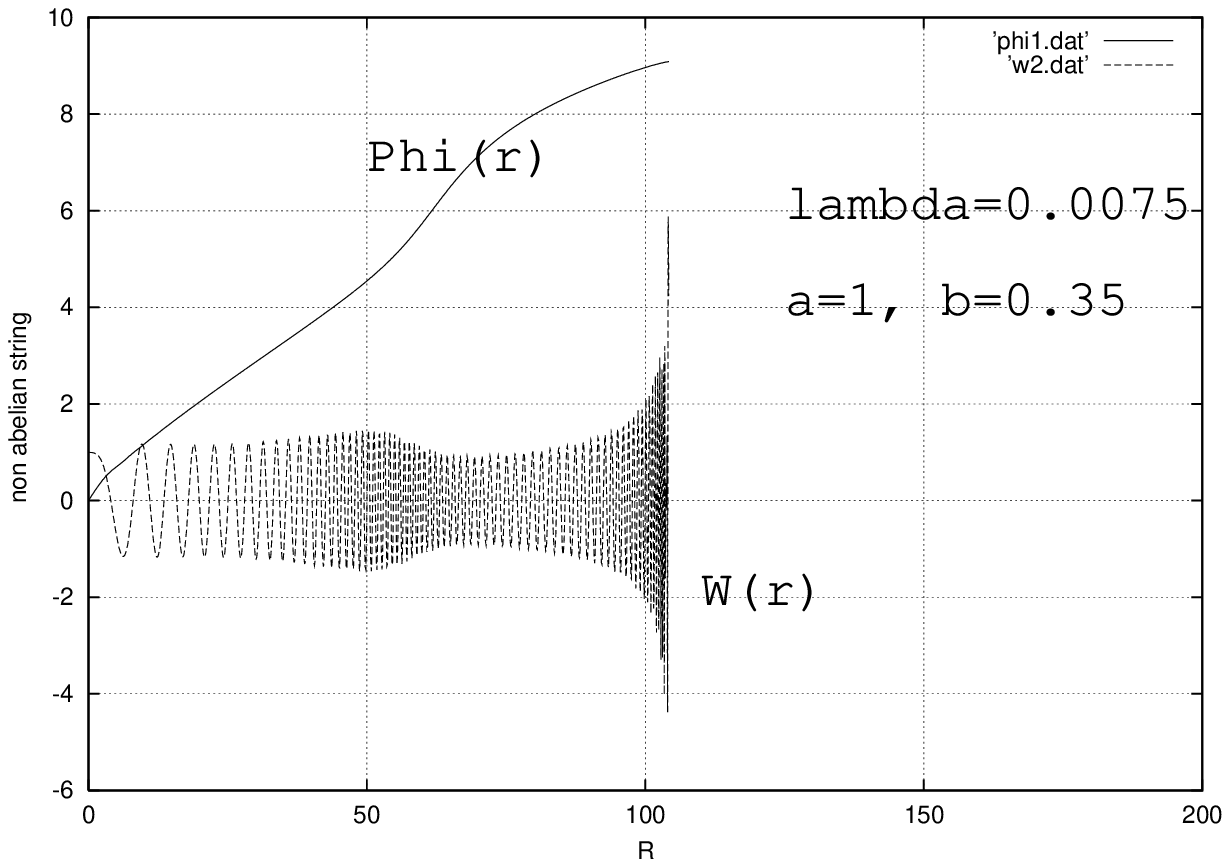}
\hfil}
\centerline
{\hfil\epsfysize=50mm\epsfbox{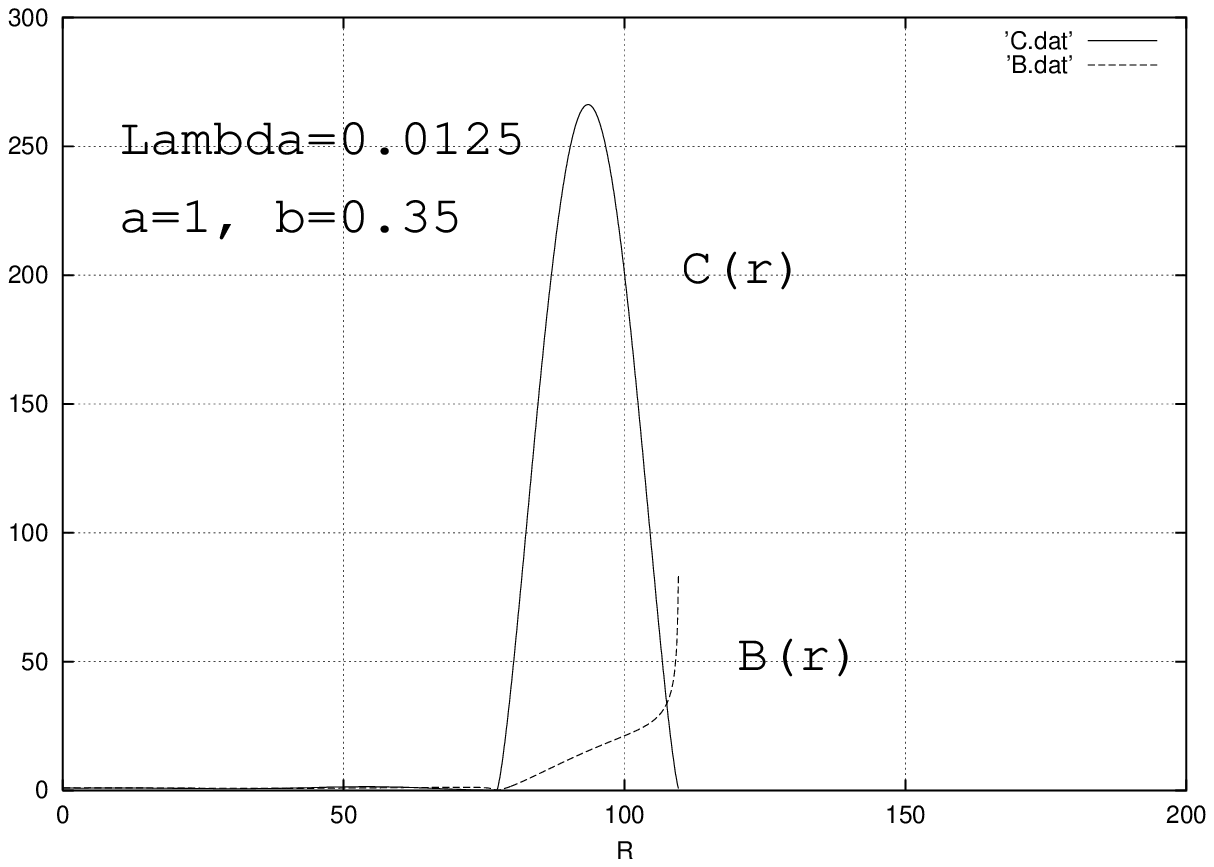}   
 \hfil\epsfysize=50mm\epsfbox{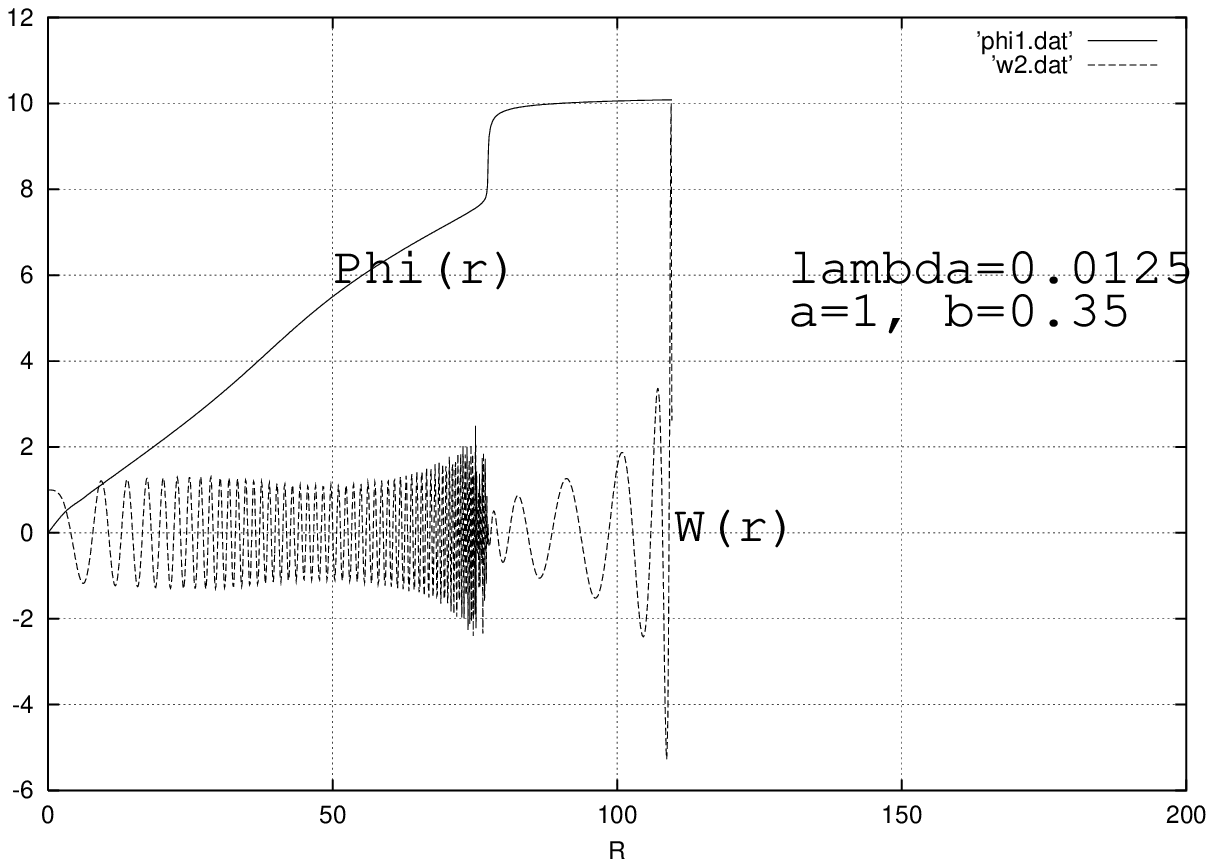}
\hfil}
\centerline
{\hfil\epsfysize=50mm\epsfbox{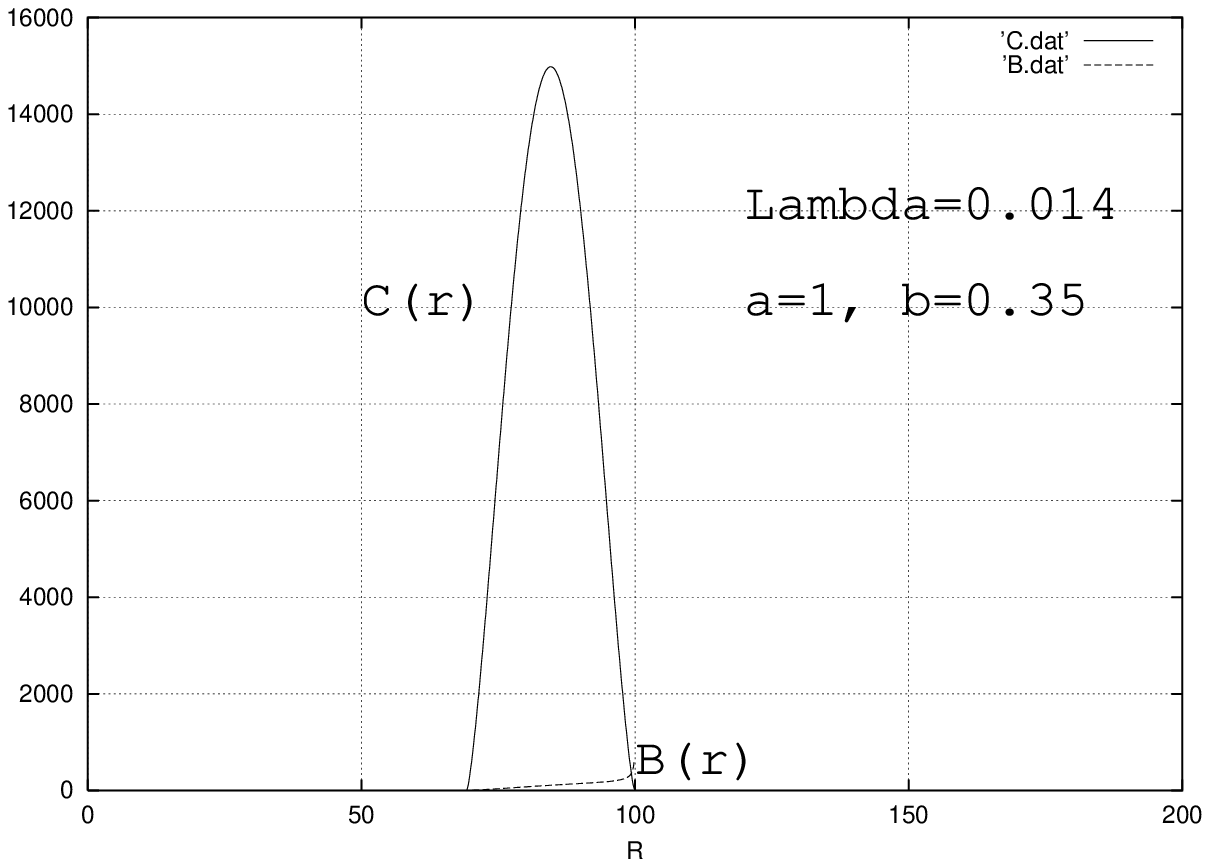}  
 \hfil\epsfysize=50mm\epsfbox{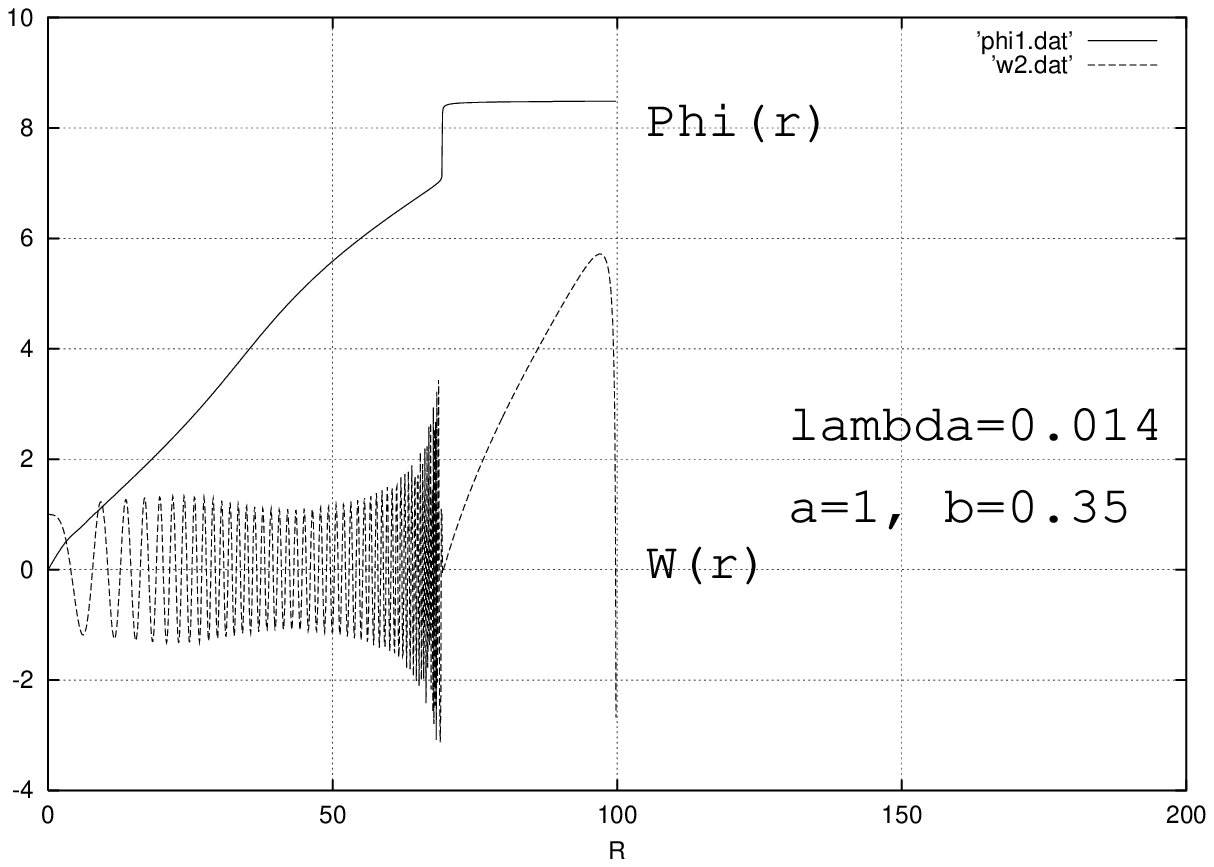}
\hfil}
\centerline
{\hfil\epsfysize=50mm\epsfbox{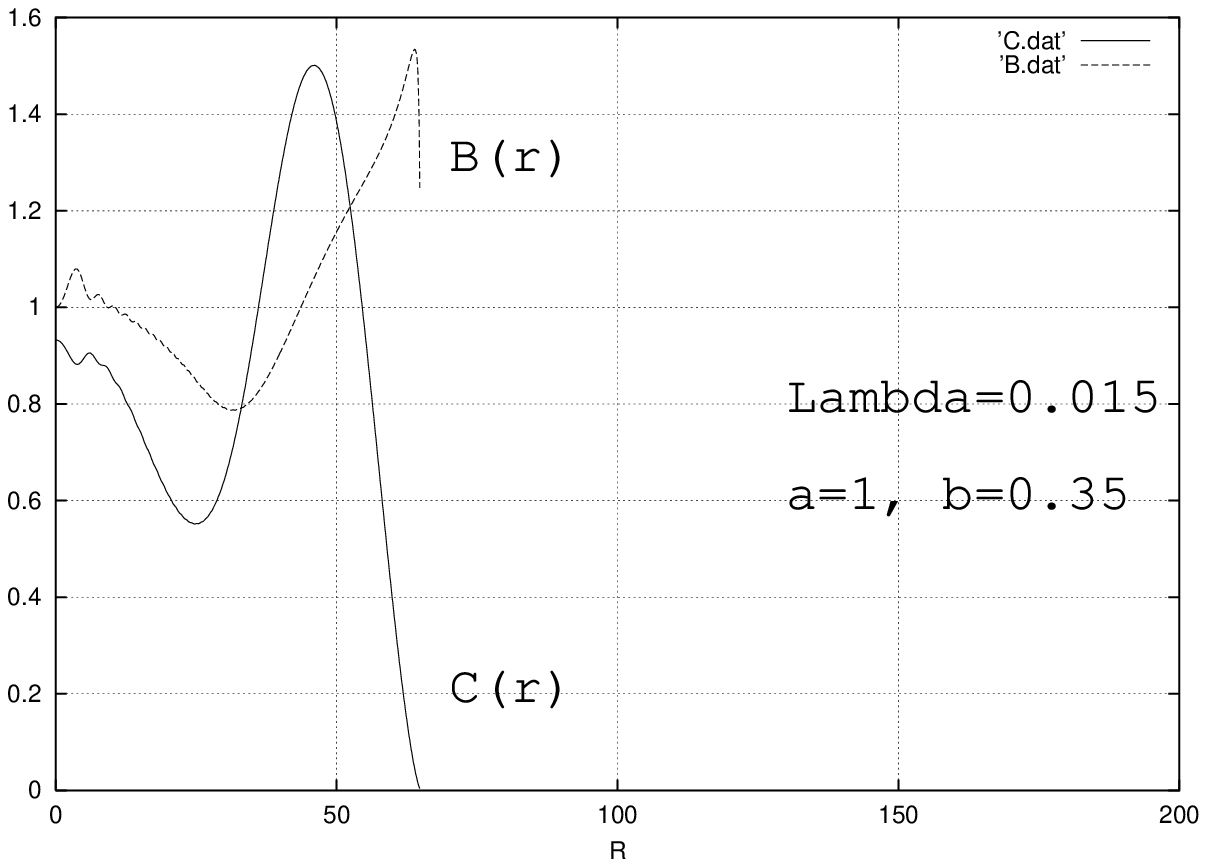}  
 \hfil\epsfysize=50mm\epsfbox{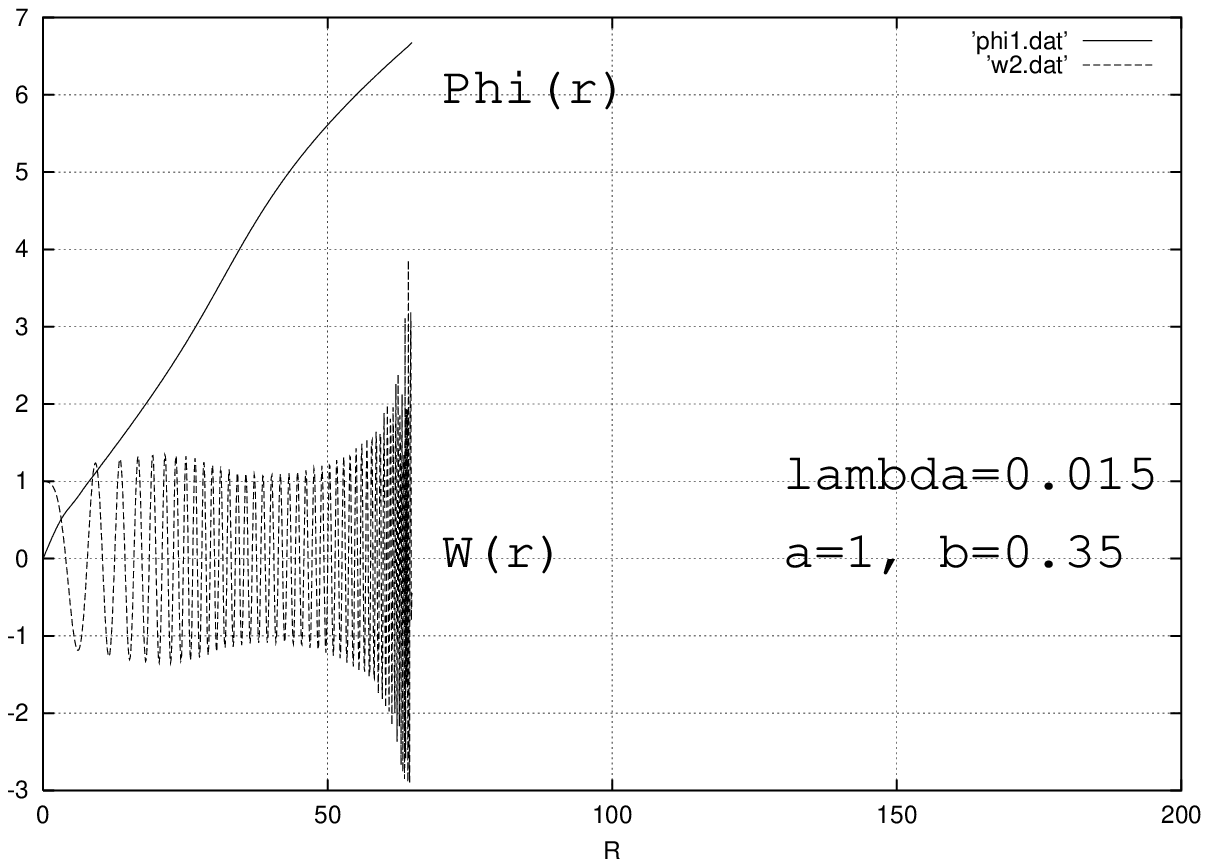}
\hfil}

Figure 2. Plot of $B(r), C(r), W(r),\Phi(r)$  for positive $\Lambda$ and initial values $ a=1, b=0.35$.
 We took  $\Lambda$ 0, 0.0005, 0.001, 0.0025, 0.0034, 0.00425, 0.0075,  0.0125, 0.014, and
 0.015  respectively. $\kappa =1$.$r_0=0.01$.

\centerline
{\hfil\epsfysize=50mm\epsfbox{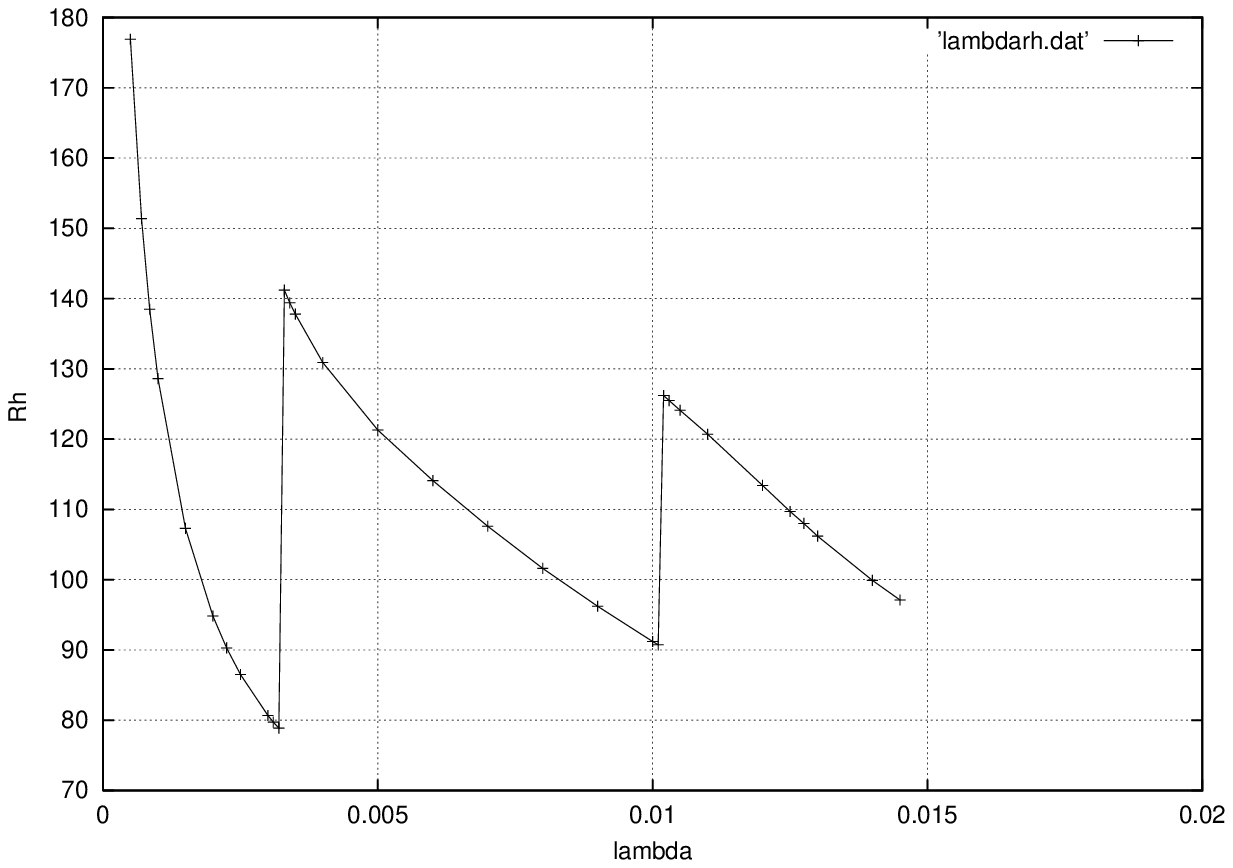}   
\hfil}
Figure 3. Plot of the horizon $r_h$ versus $\Lambda$, for $r_0=0.01$.

\centerline
{\hfil\epsfysize=50mm\epsfbox{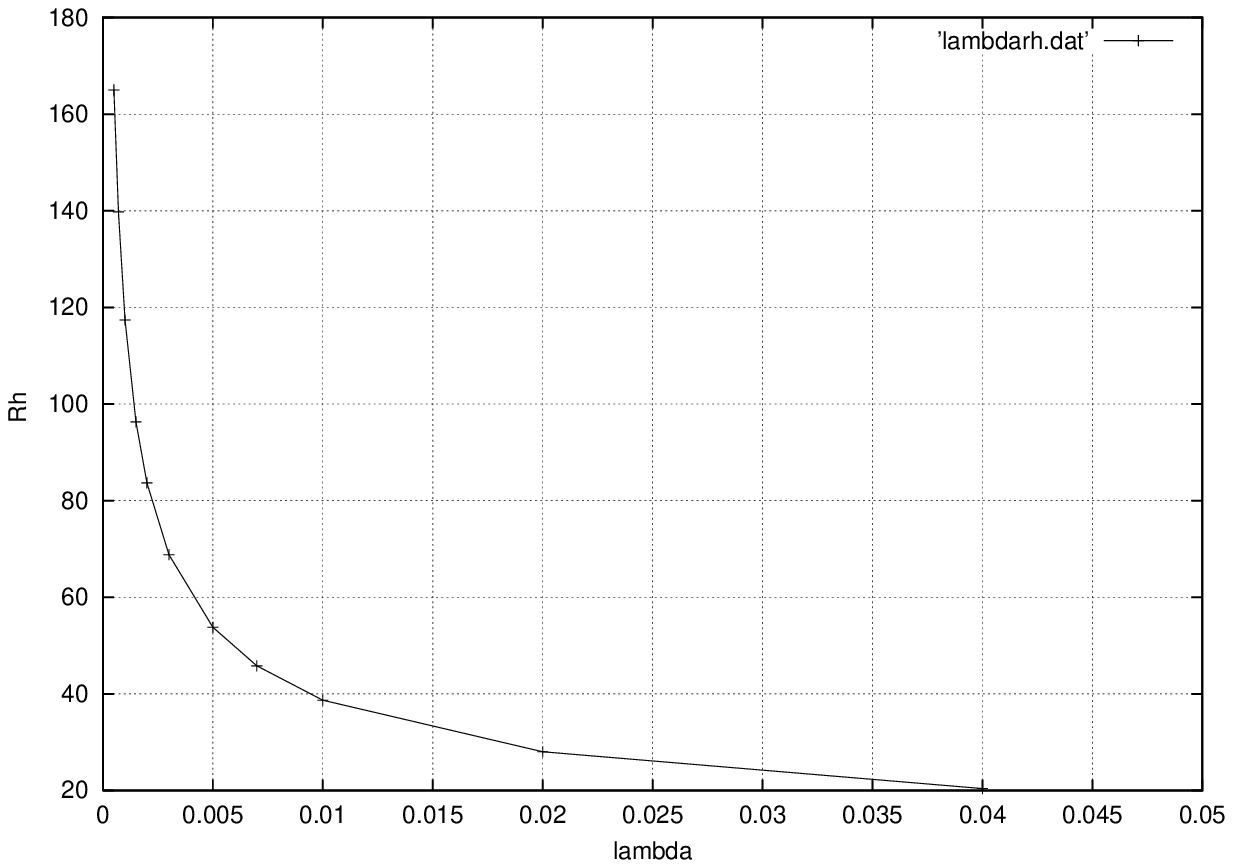}   
\hfil}
Figure 4. Plot of the horizon $r_h$ versus $\Lambda$, for $r_0=10^{-8}$.

\section{Conclusions}

This paper is a first attempt to investigate numerically the dyon solutions of the EYM model with a cosmological
constant on a cylindrically symmetric space time.
In the numerical code, we start with  regular initial values of the field variables at a small value of the core radius of the string
 and varied $\Lambda$ continuously positive as well as negative.

In contrast with the spherical symmetric EYM solutions with negative cosmological constant, our numerical
solutions seems to be regular every where without any a priori assumption on the asymptotic behaviour. The solution behaves 
like a cosmic string. However further investigation is needed.
For positive cosmological constant, the numerical solutions are quite different from the spherically symmetric counterpart 
models. First of all, the rapid oscillatory behaviour of $W$ is not found in earlier studies. The classification on the number of
nodes of $W$ makes no sense in our solution. Secondly, The position of $r_h$ moves towards the origin
for increasing $\Lambda$, but shows a periodic behaviour when $\Lambda$ is further increased.
When the core radius is decreased, this periodic behaviour dissapears.
The solution found here needs further investigation, specially for different initial value. Further, the charges should be
incorporated in the code, for example, the electric
\begin{equation}
Q_E=\frac{1}{4\pi g}\Bigl(\int r\Phi W\tau_\varphi dr -r \int \Phi'\tau_r d\varphi \Bigr)
\end{equation}


\end{document}